\documentclass[twocolumn,floatfix,nobalancelastpage,nofootinbib]{revtex4}

\usepackage{amsmath}
\usepackage{amssymb}

\usepackage{graphicx}
\usepackage{subfigure}

\usepackage{algorithm}
\usepackage{algorithmic}

\begin{document}

\title{Multi-Agent Reinforcement Learning and Genetic Policy Sharing}
\author{Jake Ellowitz}
\email{jellowitz@clarku.edu}
\date{\today}
\affiliation{Clark University}

\begin{abstract}
\begin{center}
\textbf{Abstract}
\end{center}
The effects of policy sharing between agents in a multi-agent dynamical system has not been studied extensively. I simulate a system of agents optimizing the same task using reinforcement learning, to study the effects of different population densities and policy sharing. 
I demonstrate that sharing policies decreases the time to reach asymptotic behavior, and results in improved asymptotic behavior.
\end{abstract}

\maketitle

\section{Introduction}
Human society can be thought of as a system of many interacting intelligent agents in which learning is key. Learning completely independently is a very ineffective facilitation of knowledge, which is why people share information with each other, allowing humans to better accomplish their tasks and goals. Enhancements in task performance through communication is observed in honeybee colonies~\cite{sherman} as well as in bacterial colonies~\cite{ben-jacob}, which too can be modeled as systems of agents in a reinforcement learning system. It is clear that sharing information can improve the performance of multi-agent learning systems. 

Sharing information has been shown to speed up task optimization in a reinforcement learning based multi-agent simulation, though not have an effect on the asymptotic performance of the learning task~\cite{tan,panait}. This effect is not necessarily general in multi-agent intelligent systems. There are different types of information which can be shared between agents, in this paper we will only consider \emph{policy assimilation}--when one agent absorbs the superior policy of another agent. 

In order to investigate the effects of sharing information on the optimization time and the asymptotic behavior of the system, I implement a robust learning algorithm to ensure that the learning algorithm can keep up with the evolution of the environment. Following such I investigate the effects of changing the population density of the system, as well as the probability in which the agents share information. We will see that the system displays behavior which displays a dependence on information sharing in its asymptotic behavior. 

Often times the multi-agent reinforcement learning problem is studied by applying methods in single-agent reinforcement learning. We can use the presented results to understand how single-agent reinforcement learning can be extended and improved to multi-agent reinforcement learning through sharing information. The results also display a new type of effect on asymptotic behavior, more specifically that sharing can in fact change the asymptotic behavior of intelligent multi-agent systems.

\subsection{Reinforcement Learning Overview}
I will briefly cover reinforcement learning, and go in depth only in the primary algorithm used in the simulations. Information about other algorithms mentioned  as well as all other aspects of this tutorial can be found elaborated in Ref.~\cite{sutton,mh_phd}. 

Reinforcement (RL) learning is a subfield of machine learning concerned with finding the best set of \emph{actions} for an agent in an environment such that its long term \emph{reward} (from the environment) is maximized. The agent is the learner and decision maker. The environment is what the agent interacts with; it is everything out of the agent's immediate control. Each time the agent takes an action, it is presented with a new situation by the environment. We call this situation the agent's \emph{state}. Though trial and error, the agent gradually discovers the best set of actions to take in certain states.

\begin{figure}[ht]
    \centering
    \includegraphics[scale=0.5]{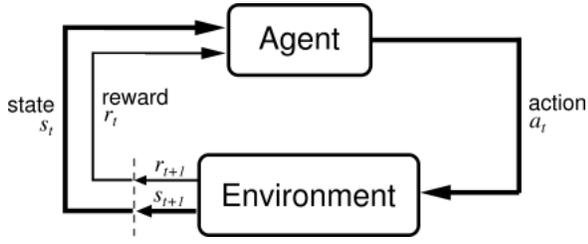}
    \caption{\label{fig:sutton_flow}A RL flowchart~\cite{sutton}}
  \end{figure}

The agent and the system interact in a sequence of discrete time steps. At each time step $t$, the agent receives a representation of the environment's state $s_t\in\mathcal S$, where $\mathcal S$ is the set of all possible states. Do not confuse the \emph{entire system} with the \emph{environment}, the environment is merely the agent's local observation. Nevertheless, the agent then takes an action $a_t \in \mathcal A(s_t)$, $\mathcal A(s_t)$ being the set of all possible actions in state $s_t$. At the next time step $t+1$, the agent receives a reward $r_{t+1}$ from the environment, as well as its new state $s_{t+1}$. See figure~\ref{fig:sutton_flow}.

The agent's decisions are governed by its \emph{policy} which maps \emph{states} to \emph{actions}:
  \[
  \pi\,:\mathcal S\rightarrow \mathcal A(s)
  \]
The agent aims to find the policy which maximizes its rewards.

\subsection{Key Aspects of Reinforcement Learning}
The algorithms in RL are designed with the assumption that the system composed of the agent and its environment is Markovian, or that it has the Markov Property. In principle, there is no reason why it cannot contain any other types of information (for instance, memory). In RL, the primary focus is on the decision making process, not designing the state signal. Accordingly, we want the state signal to be compact, which is why we have it reflect immediate sensory information. Though a system may not be Markovian, at least an approximation to a Markovian system is good enough in order for the RL algorithms to work properly.

The \emph{policy} $\pi(s,a)$ is a set of probability distributions specifying the probability the agent will take a certain action a given its states. This probability is given as
  \[
  \pi(s,a) = p\{a'=a|s\},
  \]
and
  \[
  \sum_{a'}\pi(s,a') = 1.
  \]
The \emph{return} at step $t$ $R_t$ is defined
  \[
  R_t = \sum_{k=0}^{T} \gamma^k r_{t+k+1},
  \]
where $T$ is the \emph{episode duration} (the number of time steps the system takes to reach a terminal state, meaning when the system has achieved an intermediate or final goal) and $\gamma \in [0,1]$ is the \emph{discount parameter}. The discount parameter specifies present consideration of past events during an episode.

The \emph{value function} for a policy $\pi$ is given by
  \[
  V^\pi(s) = E_\pi\left.\left\{R_t\right|s_t = s \right\}.
  \]
The value function tells us our expected returns given a state. It will tell us how good that state is: a state with a higher value function is more preferable because we expect a higher long-term reward. Similar to the value function, we define the \emph{action-value function} for a policy $\pi$ as
  \[
  Q^\pi (s,a) = E_\pi\left.\left\{ R_t\right| s_t=s, a_t = a\right\}.
  \]
The action-value function tells us how good it is to take a certain action in a certain state using the same logic as in the value function. It is easy to see that
  \[
  V^\pi(s) = \sum_{a'\in\mathcal A(s)} Q^\pi(s,a').
  \]
Optimizing performance in a reinforcement learning system corresponds to maximizing our value and action-value functions. Accordingly, we aim to find the \emph{optimal value function} and the \emph{optimal action-value function} $V^*$ and $Q^*$ by
  \begin{align*}
  V^*(s) &= \max_\pi V^\pi (s), \\
  Q^*(s,a) &= \max_\pi Q^\pi (s,a),
  \end{align*}
for all $s \in \mathcal S$, $a\in \mathcal A(s)$. Thus we need some method of determining which policies are better than others. We do this by comparing value functions:
  \[
  \pi \ge \pi' \Leftrightarrow V^\pi(s) \ge V^{\pi'}(s)
  \]
for all $s\in\mathcal S$. Following this, we can define the \emph{optimal policy} $\pi^*$ as
  \[
  \pi^* \ge \pi
  \]
for all $\pi$. Note $\pi^*$ is not necessarily unique.

\subsection{Policy Improvement}
The method involved in finding the optimal policy is fairly straightforward and applies generally to different methods of optimization. The steps are:
  \[
  \pi_0 \xrightarrow{E} V^{\pi_0},Q^{\pi_0} \xrightarrow{I} \pi_1 \xrightarrow{E} \cdots \xrightarrow{I}\pi* \xrightarrow{E}V^*,Q^*,
  \]
with $\xrightarrow{E}$ and $\xrightarrow{I}$ denoting policy evaluation and policy improvement, respectively. The methods of policy improvement relevant here are \emph{$\varepsilon$-greedy}
  \[
  \pi(s,a) = \left\{
    \begin{array}{lr}
      1-\varepsilon+\varepsilon/|\mathcal A(s)| & 
        \text{ if $a=\arg\max_{a'} Q(s,a')$} \\
      \varepsilon/|\mathcal A(s)| & 
        \text{ if $a\ne\arg\max_{a'} Q(s,a')$} \\
    \end{array}
  \right.,
  \]
and \emph{softmax}
  \[
  \pi(s,a) = \frac{e^{Q(s,a)/\tau}}{\sum_{a'}e^{Q(s,a')/\tau}},
  \]
where the \emph{temperature parameter} $\tau$ specifies the randomness of decisions.

\subsection{Q-Learning}

In Q-Learning (QL) we approximate the action-value function according to
  \begin{multline}\label{eq:ql}
  Q(s_t,a_t) := Q(s_t,a_t) + \\
    \alpha\left\{
      r_{t+1}+\gamma\max_{a'}Q(s_{t+1},a') -Q(s_t,a_t)
    \right\},
  \end{multline}
with $\alpha$ being the \emph{learning rate} and $\gamma$ (as before) the discount parameter~\cite{watkins_phd}. Just like the discount parameter, we want $\alpha\in[0,1]$. In order for a policy evaluated by QL to converge in a discrete number of time steps, it is proven in Ref.~\cite{watkins} that
  \begin{equation}\label{eq:ql_conv}
    \begin{split}
    \sum_{k=1}^{\infty} \alpha_k &= \infty, \\
    \sum_{k=1}^{\infty} \alpha_k^2 &< \infty.
    \end{split}
  \end{equation}
In order for the criterion in equation~\eqref{eq:ql_conv} to be met, we define
  \[
  \alpha_k (s,a) = \frac{1}{k(s,a)},
  \]
where $k(s,a)$ is the number of times the state-action pair $s,a$ has been visited. I decided to use this method to ensure the accuracy of the approximation of the value and action-value functions, rather than methods in finding the most accurate constant-$\alpha$ value. For an approach to the latter method, see Ref.~\cite{eyal}. An implementation of QL can be found in algorithm~\ref{alg:ql}.

  \begin{center}
    \begin{scriptsize}
    \begin{algorithm}
    \caption{\label{alg:ql}Q-Learning}
    \begin{algorithmic}
    \WHILE {True}
      \STATE $a_{t} := \pi(s_{t})$
      \STATE Observe $r_{t+1}, s_{t+1}$
      \STATE $\alpha(s_t, a_t) := \alpha(s_t,a_t) / (1+\alpha(s_t,a_t))$
      \STATE 
        $Q(s_t,a_t) := Q(s_t,a_t) +$
      \STATE
        \hfill$\alpha(s_t, a_t)\left\{
        r_{t+1} + \gamma\max_{a} Q(s_{t+1},a) - Q(s_t,a_t)
        \right\}$
      
      \STATE Iteration := Iteration + 1
      \IF {Iteration $>$ Max Iteration \textbf{or} Terminal State}
        \STATE Update $\pi$ ($\varepsilon$-greedy or softmax)
        \STATE Iteration := 0
      \ENDIF
    \ENDWHILE
    \end{algorithmic}
    \end{algorithm}
    \end{scriptsize}
  \end{center}

\section{The System}
\begin{figure}[ht]
    \centering
    \includegraphics[scale=0.5]{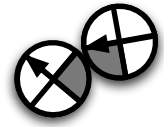}
    \caption{An illustration of two touching agents. The one to the left is in state 4, whereas the other agent is in state 1. If one of these agents broadcasts its policy, the other would receive the broadcast.}
    \label{fig:bot_state}
  \end{figure}

The system I investigate is an enclosed two-dimensional square arena containing circular bot-agents. The density $\rho$ corresponds to the surface area of the bots divided by the total area of the arena. Each bot has sensors covering $N$ equal wedges of its circumference. The sensors are numbered $0,1,\cdots,N-1$. In the simulation, we fix $N=4$. If there is some object touching the agent at some angle $\phi$ from the agent's orientation, the $n^{\rm th}$ sensor is activated, $n$ given by
  \[
  n = \left\lfloor \frac{N\phi}{2\pi} \right\rfloor.
  \]
For each state, all of the other $N-1$ sensors can be \emph{on} or \emph{off}. Thus, we can define our state $s$ as
  \[
  s = \sum_{n=0}^{N-1} \mu_n 2^n,
  \]
with $\mu_n=1$ if the $n^{\rm th}$ sensor is activated, and $\mu_n=0$ if it is not. According to this representation, there will be $2^N$ possible states:
  \[
  \mathcal S = \{0,1,\cdots,2^N-1\}.
  \]
An illustration can be found in figure~\ref{fig:bot_state}. The bots are smooth, hard disks, which collide with the walls and other bots completely inelastically. The simulation is initialized by placing all of the agents in the arena, all at different random initial orientations $\theta_{0,i}$, such that none overlap.
  
Each bot-agent has a fixed set of actions $\mathcal A$ for all $s\in \mathcal S$--either rotate to one of its other fixed, evenly spaced angles\footnote{If the bot-agent is oriented to angle $\theta$, this corresponds to changing this angle to $\theta+2n\pi/N$ for $n\in\{1,2,\cdots,N-1\}$.}, or move forward in a straight line at a universal constant speed along the orientation $\theta$ of the bot-agent. Thus $|\mathcal A(s)| = N$ for all $s\in\mathcal S$. Respectively, the next learning iteration occurs when the agent reaches its terminal orientation (it is done rotating towards its new angle), or it experiences a collision. These occurrences are called \emph{events}. After an event occurs, the agent broadcasts its policy with probability $p$ to all other adjacent agents touching the original agent's circumference (as in figure~\ref{fig:bot_state}). Algorithm~\ref{alg:pi_share} displays how the policy sharing works, with $\mathcal B$ being set of bot-agents touching bot-agent $A$ when $A$ broadcasts its policy. This policy sharing algorithm is genetic in that it evaluates the fitness of a policy based on its value function, and proceeds with na\"{\i}ve evolutionary selection.

  \begin{algorithm}[h]
    \caption{Policy Sharing Algorithm}
    \label{alg:pi_share}
    \begin{algorithmic}
      \IF {Random $<p$}
        \FOR {$B\in\mathcal B$}
          \IF {$\pi^A \ge \pi^B$}
            \STATE $V^B(s) := V^A(s) \forall s$
            \STATE $Q^B(s,a) := Q^A(s,a) \forall s,a$
            \STATE $\pi^B := \pi^A$
          \ENDIF
        \ENDFOR
      \ENDIF
    \end{algorithmic}
  \end{algorithm}

Do not confuse the traditional notion of state with that used in RL: states are measured in discrete intervals during each time step, or learning iteration. When the agent is moving there is no \emph{moving state} associated with it. The agent's next state is determined by the readings from its sensors after an event. 

\begin{figure}[t]
    \centering
    \subfigure[Small arena: $150\times150$, $\rho=0.279$\label{fig:arena_pic_small}]{\includegraphics[scale=0.5]{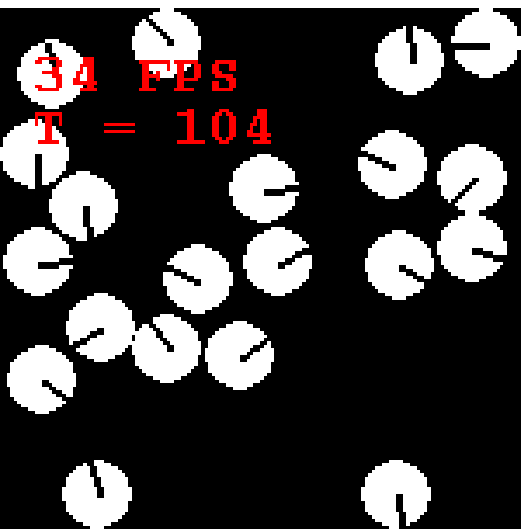}}
    \subfigure[Large arena: $200\times200$, $\rho=0.0236$\label{fig:arena_pic_large}]{\includegraphics[scale=0.5]{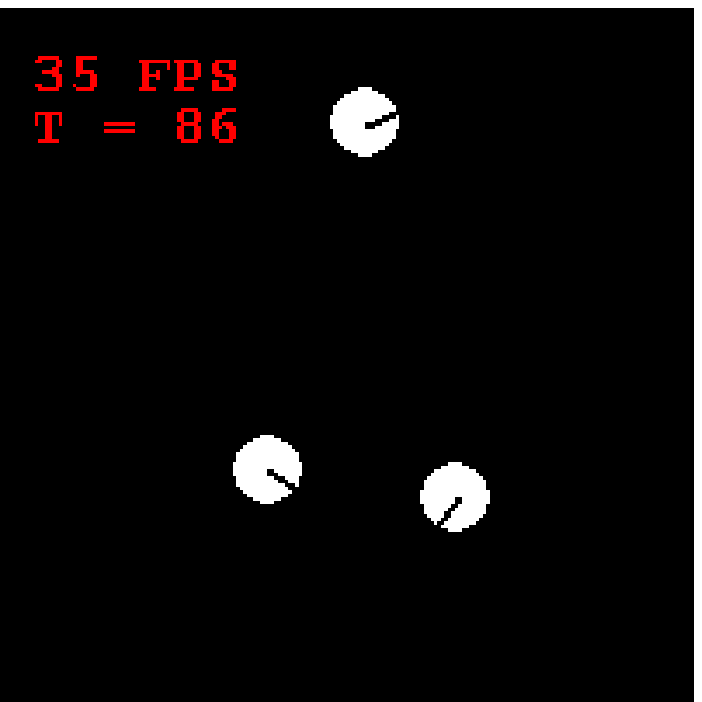}}
    \caption{Images of the small and large arenas used in the simulations.}
    \label{fig:arena_pic}
  \end{figure}

The simulations are run on two square arenas. If $L$ is the length of a side of the arena and $R$ is the bot-agent radius, the large arena has $L/R=20$, whereas the small arena has $L/R=15$. The large arena holds up to 50 agents, and the small arena holds up to 25 agents. Some images of these different setups are found in figure~\ref{fig:arena_pic}.

Each agent's task is to travel the greatest distance. We represent this task by specifying the reward the agent will receive after each action:
  \[
  r_{t+1} = -C + kD_{a_t},
  \]
with $C, k\in \mathbb R^+$, and $D_{a_t}$ is the distance traveled as a result of action $a_t$. The $C$ parameter is to discourage actions in which $D$ is small, and the $k$ parameter is to make sure that the rewards from traveling do not drown out $C$. Good behavior corresponds to short convergence times (quick learning) and a higher average speed (longer distance per unit time). 

\subsection{Focusing on One Learning Algorithm}
\begin{figure}[h!]
    \centering
    \includegraphics[scale=0.5,angle=270]{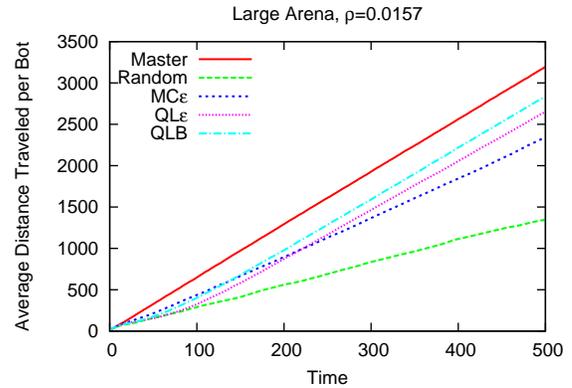}
    \caption{A comparison of learning algorithms for a low arena density. The $\varepsilon$ corresponds to using an $\varepsilon$-greedy policy evaluation, whereas the B corresponds to a softmax policy evaluation. There is no policy sharing.}
    \label{fig:all_comp}
  \end{figure}

I conduct the primary simulations using a single learning algorithm. Though interesting behavior is not limited to the best algorithm (the one which achieves the objective to the greatest extent), we choose one to reduce the number of variables. I decided to choose the best-performing algorithm. This is because as the sharing and density parameters are varied, I want to be safe and not have buried behavior contained in the algorithm as it struggles to optimize a policy. A typical comparison of learning algorithms considered is displayed in figure~\ref{fig:all_comp}. The behavior displayed is observed across different densities. It is clear that the softmax QL algorithm is quick in adapting to its environment. All simulations have $\gamma=0.9$ and $\tau=0.5$.

\section{Results}
\begin{figure}[t]
    \centering
    \includegraphics[scale=0.5]{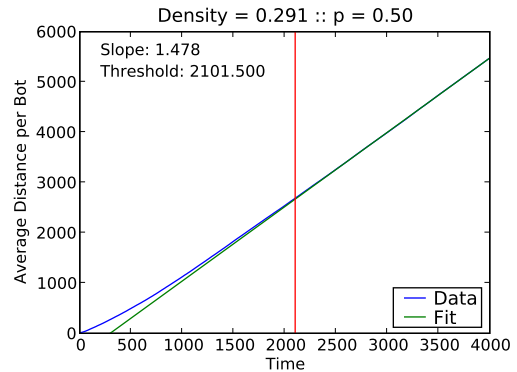}
    \caption{The vertical red line specifies the point in time when the system converged. The sharing probability is $1/2$ in this case.}
    \label{fig:threshold}
  \end{figure}

As we have already seen, the QL algorithm converges. This convergence shows itself as the constant average velocity per bot displayed in figure~\ref{fig:all_comp}, as the agents' policies become static. This is the criterion we will use to determine when the system converges--or when the average distance per bot becomes linear in time. That point is found by fitting to the tail of the curve, and extrapolating the fit line backwards in time. The point where the fit curve deviates from the data is the \emph{threshold}, or when the system displays its asymptotic behavior. See figure~\ref{fig:threshold}.

\subsection{Hypotheses}
For lower arena densities, the system is less complicated and, accordingly, the agents should experience similar situations more often and optimize more quickly. We should also see the effects of sharing more prominently in higher densities, as there will be more interactions per bot per time step. We can also expect that in general some sharing should speed up convergence, however too much sharing might slow down convergence--the system will act \emph{greedily} in the beginning, optimizing immediate rewards, impairing the efforts of exploration. Something that we cannot guess is where the fastest convergence lies along the sharing probability space. We are also unsure as to the effect of sharing probability $p$ on the asymptotic behavior, though other sources mentioned earlier lead us to believe sharing does not affect the long-time behavior. 

\subsection{Simulation Results and Discussion}

\begin{figure}[h]
    \centering
    \includegraphics[scale=0.5,angle=270]{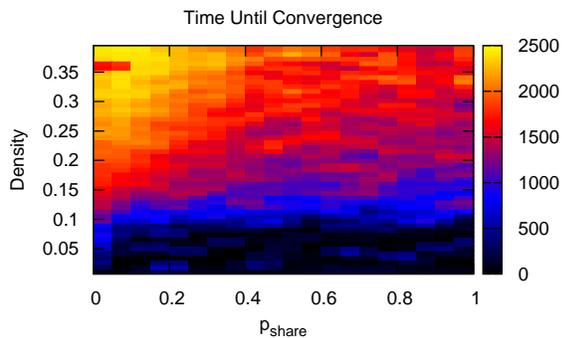}
    \caption{Simulation results for the large arena. Notice the fastest convergence times correspond to the highest sharing probabilities.}
    \label{fig:l_conv}
  \end{figure}

\begin{figure}[h]
    \centering
    \includegraphics[scale=0.5,angle=270]{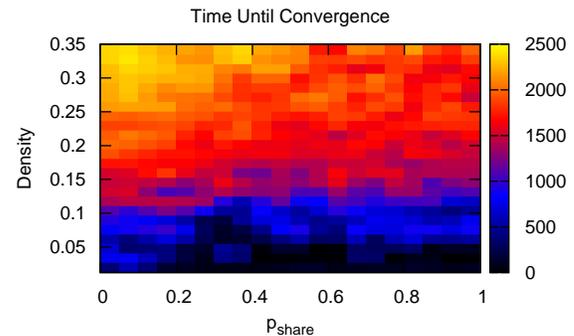}
    \caption{Simulation results for the smaller arena. Notice similar behavior to that of the large arena in figure~\ref{fig:l_conv}.}
    \label{fig:s_conv}
  \end{figure}

In figure~\ref{fig:l_conv} we see that for lower densities, the sharing does not have such a strong effect on the convergence time. Also, we observe that, in general, lower densities converge quicker. We observe similar behavior in the smaller arena, but it curiously takes \emph{more} time for the smaller arena to converge. See figure~\ref{fig:s_conv}. Regardless, the general trends seem to apply to both systems.

  \begin{figure}[h]
    \centering
    \includegraphics[scale=0.5,angle=270]{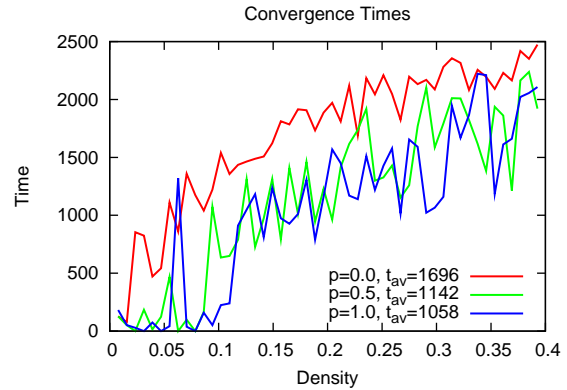}
    \caption{A closer look at some sharing probabilities from figure~\ref{fig:l_conv}. Some unfitting spikes are due to artifacts in the convergence time calculations (see figure~\ref{fig:threshold}), for instance if the data ran closely parallel to the tail linear fit for a long duration of time.}
    \label{fig:sel_conv}
  \end{figure}
 
  \begin{figure}[h]
    \centering
    \includegraphics[scale=0.5,angle=270]{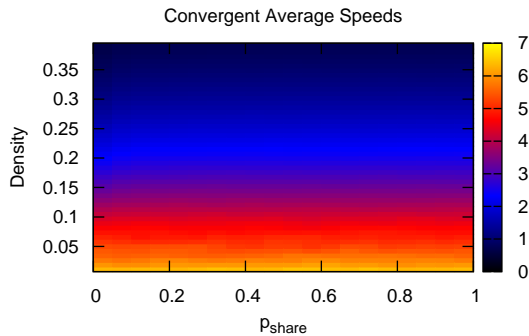}
    \caption{Simulation results for the large arena, displaying the asymptotic performance of the collective. Notice the slightly darker band for no sharing (to the left), then the rest of the behavior is uniform.}
    \label{fig:l_conv_speed}
  \end{figure}

\begin{figure}[h]
    \centering
    \includegraphics[scale=0.5,angle=270]{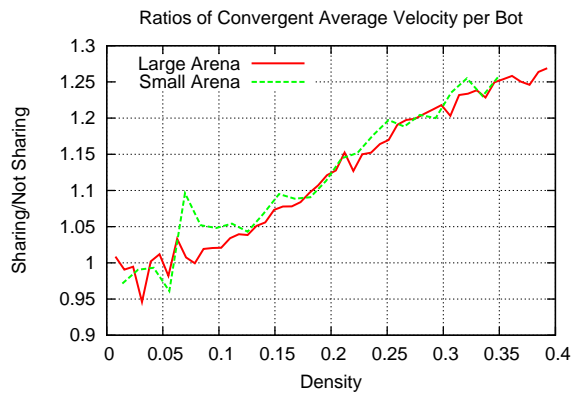}
    \caption{The ratio of the sharing terminal speeds over the independent terminal speeds as a dependency on $\rho$.} 
    \label{fig:sha_nosha_ratio}
  \end{figure}

Since we are curious about asymptotic performance, we investigate the asymptotic average velocity, as for example in figure~\ref{fig:threshold}. In figure~\ref{fig:l_conv_speed}, there appears to be \emph{no} difference between low sharing probabilities and high sharing probabilities. There \emph{does} appear to be a difference between no sharing at all (independent) and any amount of sharing. In order to get a closer look at this ratio, refer to figure~\ref{fig:sha_nosha_ratio}. We observe a heavy discrepancy from the independent (non-sharing) system in the asymptotic behavior of any sharing system as the density increases.

\begin{figure}[h!]
    \centering
    \includegraphics[scale=0.5,angle=270]{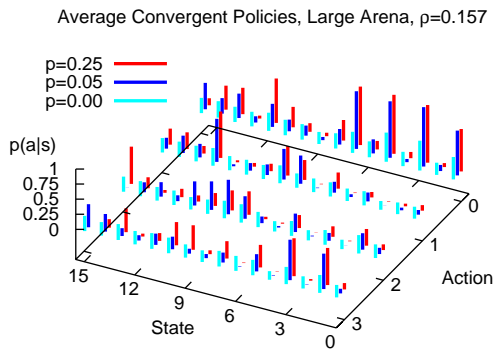}
    \caption{A low share probability $p$ and a high share probability display the collective all adapting the same preferences. With $p=0$ we see that the average policies level out, indicating agents do not coordinate.}
    \label{fig:cooperation}
  \end{figure}

When each agent is finding its own local optimal policy, it eventually will have to make some sort of preference towards how it resolves certain situations--for instance, to turn left or turn right after encountering a collision. These preferences become approximately permanent after a policy is well established. These preferences will cause any given agent in the collective to butt heads, in a way, with the other agents, as it prefers (for instance) a counterclockwise vortex versus a clockwise one. When the agents are forced to swap policies, then the collective converges to a single preference. This exact behavior is displayed in figure~\ref{fig:cooperation}. Make no mistake, the independent agents' policies \emph{do} optimize, but with independent reinforcement learning it is too difficult for the agents to coordinate.

\section{Conclusions}
We see that the time until convergence depends on the arena size, the arena density, and the sharing probability. The smaller arena experienced slower convergence times, the lower densities experienced faster convergence times, and the higher sharing probability experienced faster convergence times. The sharing probability yielding the fastest convergence time appears to be in the regime of 1, and not somewhere in between 0 and 1 as we expected. 

Perhaps the most interesting result of this simulation is in contrast to Ref.~\cite{tan,panait}, we find that sharing policies significantly affect on the asymptotic behavior of the system, especially for higher densities. As the system develops, each agent develops its own preferences towards resolving different states. These preferences become permanent after the agent has a well established policy. If there is any sharing ($p>0$), as the system runs for very long periods of time, all of the agents will adopt a uniform preference and the collective becomes coordinated. The specified task is being performed better in the sharing case than in the independent case. This improvement shows itself more prominently at greater densities, and appears to be independent of the arena size.

The genetic policy sharing used (outlined in algorithm~\ref{alg:pi_share}) is by no means the best algorithm. Variants might prove to perform better, such as averaging policies and value functions rather than erasing the past information an agent had accumulated. Regardless, these results can be explained quite plainly: I have demonstrated that the asymptotic behavior of a RL system can be improved through a policy sharing mechanism.

\subsection{Future Work}
The fluctuations of the system are not discussed here. Such aspects of the system could yield interesting behavior. For example, if the emergent cooperative behavior of the system increases or decreases the fluctuations of the performance measure (total distance traveled per agent with respect to time). 

One could also study the effects and robustness of other learning algorithms in the context of the system. In Ref.~\cite{guo}, Maozu Guo \emph{et al.} find that a learning algorithm based on simulated annealing performs very well. It would be interesting to see if a simulated annealing learning algorithm performs as well as QL, and to see how the system behaves under varying parameters with respect to the two learning algorithms. Additional studies of alternative algorithms include varying the sharing algorithm, and introducing an inhomogeneity of learning algorithms. Furthermore, one can introduce an inhomogeneity of agents by, for example, varying the agent radius.

A mediating algorithm can be introduced to coordinate the agents. This algorithm can be based on RL or some other subfield of machine learning. The effects of this meta-algorithm might introduce some very interesting behavior.

The nature of the system lends itself well to a spacial diffusion simulation. This can be done by increasing the arena size, and placing many agents in a dense group at the arena center. In addition to studying spacial diffusion, one can study the diffusion of knowledge among the collective. One way this can be done by placing a master agent among dummy agents, and tracking how the knowledge diffuses, and how that diffusion affects the collective's task performance.

\begin{acknowledgments}
Thanks go to Benny Brown for helping me get this project off the ground and also to Dr.\ Jim Crutchfield. I would also like to thank Dr.\ Ping Xuan for a helpful discussion. The research was conducted as part of the Network Dynamics Program at the Complexity Sciences Center at the University of California, Davis, and was supported by the NSF REU Program 2008 at the University of California, Davis.  
\end{acknowledgments}

\appendix
\section{Running the Simulation}
In order to run the simulation, execute the sim.py file in the primary directory. Some help is included, as well as a readme file. Contact me with any questions regarding the simulation output or the underlying processes. The simulation source code is available at 
{\begin{scriptsize}\tt http://students.clarku.edu/$\sim$jellowitz/files/release.tar.gz\end{scriptsize}}.

\bibliographystyle{apsrev}
\bibliography{main.bib}
\end{document}